\documentclass[aps,prl,amsmath,nofootinbib,amssymb,twocolumn]{revtex4-1}

\usepackage{graphicx}
\usepackage{dcolumn}
\usepackage{footmisc}

\usepackage{bm}
\usepackage{hyperref}
\usepackage{xcolor}
\usepackage{float}
\begin{document}
\setlength{\abovedisplayskip}{5pt}
\setlength{\belowdisplayskip}{5pt}
\setlength{\abovedisplayshortskip}{5pt}
\setlength{\belowdisplayshortskip}{5pt}

\preprint{}

\title{Diffuse Boosted Cosmic Neutrino Background}

\author{Gonzalo Herrera, Shunsaku Horiuchi, Xiaolin Qi}
\affiliation{Center for Neutrino Physics, Department of Physics, Virginia Tech, Blacksburg, VA 24061, USA}
\affiliation{Kavli IPMU (WPI), UTIAS, The University of Tokyo, Kashiwa, Chiba 277-8583, Japan}

\begin{abstract}
Energetic cosmic rays scatter off the cosmic neutrino background throughout the history of the Universe, yielding a diffuse flux of cosmic relic neutrinos boosted to high energies. We calculate this flux under different assumptions of the cosmic-ray flux spectral slope and redshift evolution. The non-observation of the diffuse flux of boosted relic neutrinos with current high-energy neutrino experiments already excludes an average cosmic neutrino background overdensity larger than $\sim 10^{4}$ over cosmological distances. We discuss the future detectability of the diffuse flux of boosted relic neutrinos in light of neutrino overdensity estimates and cosmogenic neutrino backgrounds. 
\end{abstract}

\maketitle

\section{Introduction}

The cosmic neutrino background (C$\nu$B) fills the Universe with a present-day average number density of $n_{\nu} \sim 336$ cm$^{-3}$ for all neutrino flavors \cite{Dolgov:1997mb, Mangano:2005cc}. Also called relic neutrinos, its flux is larger by orders of magnitude than the fluxes of any other astrophysical or terrestrial neutrino source \cite{Vitagliano:2019yzm}. However, the energy of relic neutrinos is lower than the largest neutrino mass, which is constrained by the KATRIN experiment for electron antineutrinos to $m_{\nu} < 0.8$ eV \cite{KATRIN:2021uub}, making its detection challenging. Indeed, relic neutrinos hardly induce detectable signals at Earth-based detectors since the energy depositions are very tiny, of order $\sim$ meV$-$eV, and their interaction cross section is suppressed in the Standard Model.

An intriguing idea to detect the elusive C$\nu$B is to consider cosmic rays that scatter them off, yielding a flux on Earth of relic neutrinos with larger ``boosted'' energies. Indeed, this was hinted in the 1980s when the neutrino mass was not constrained and the cosmic-ray flux was uncertain \cite{hara_sato,Hara:1980mz}. Recently, it has been pointed out that cosmic-ray scatterings off the C$\nu$B can yield a flux of neutrinos on Earth that rules out a C$\nu$B overdensity in the Milky Way of $\sim 10^{13}$, or of $\sim 10^{10}$ in TXS 0506+056, from a non-observation of an excess of events in high-energy neutrino experiments \cite{Ciscar-Monsalvatje:2024tvm}. Similar bounds have been found considering the non-attenuation of the high energy neutrino flux from NGC 1068 due to scatterings off the C$\nu$B during propagation \cite{Franklin:2024enc}. Comparable bounds were also found recently from the boosted C$\nu$B flux from cosmic-ray reservoirs \cite{DeMarchi:2024zer} accounting for detailed modelling of the neutrino-nucleus cross section at ultra high energies and cosmic-ray composition. However, the large neutrino overdensities constrained in these studies \cite{Ciscar-Monsalvatje:2024tvm, Franklin:2024enc, DeMarchi:2024zer} are unlikely to be realized in Nature, mainly due to Pauli exclusion principle which disfavours relic neutrino overdensities with values larger than $\sim 10^6$ \cite{Bauer:2022lri}.

\begin{figure}[H]
    \centering
    \includegraphics[width=0.5\textwidth]{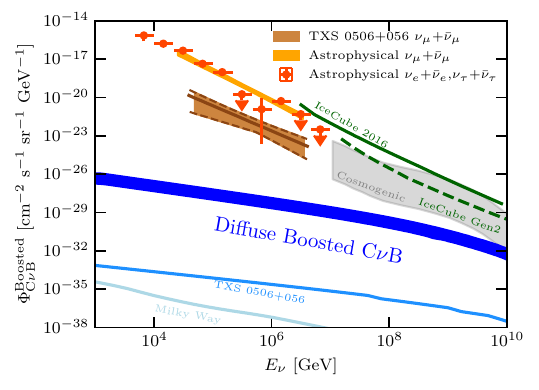}
    \caption{The flavor averaged ($\nu_\alpha+\bar\nu_\alpha$) diffuse flux of the C$\nu$B boosted by cosmic-ray protons throughout the Universe history (blue, as labeled). The thickness of the blue band captures the uncertainty in the cosmic-ray spectral index and redshift evolution of sources (see main text for details). Here, the neutrino mass is taken to be $m_\nu=0.1$ eV. For comparison, we show the boosted C$\nu$B fluxes from a consideration of Milky Way cosmic rays and from TXS 05056+056 \cite{Ciscar-Monsalvatje:2024tvm}. Also for comparison, we show the total flux of atmospheric neutrinos and diffuse extragalactic neutrinos measured by experiments at Earth \cite{Super-Kamiokande:2015qek, IceCube:2016umi, IceCube:2020acn}, as well as current upper limits of very high-energy neutrinos from IceCube (green, \cite{IceCube:2016uab}). We also show sensitivity projections for IceCube-Gen2 radio (10 yr) \cite{IceCube-Gen2:2021rkf}. In grey, we show the cosmogenic neutrino flux \cite{Fang:2017zjf}, and vary the normalization to account for uncertainty following Ref.~\cite{GRAND:2018iaj}.}
    \label{fig:boosted_flux}
\end{figure}

The calculation of Ref.~\cite{Ciscar-Monsalvatje:2024tvm}, while pioneering, was restricted to only single galaxies: the Milky Way and TXS 0506+056. Nevertheless, there must be a contribution to the diffuse flux of boosted relic neutrinos reaching the Earth that arise from all other galaxies in the Universe. Here, we calculate this diffuse flux, and find it enhances the boosted neutrino flux by many orders of magnitude, as show in Fig.~\ref{fig:boosted_flux}. By comparing the predicted boosted flux to upper limits on searches for high-energy cosmogenic neutrinos by, e.g. IceCube \cite{IceCube:2016uab}, we are able to set a world-leading limit on the C$\nu$B overdensity at the level of $\sim 10^4$, and a projected limit of $\sim 10^3$, for a neutrino with mass $m_{\nu}=0.1$ eV. Importantly, this is the first bound on the C$\nu$B over cosmological timescales/distances, and future experiments may be able to probe C$\nu$B overdensities below the critical density of the Universe.

\section{Diffuse boosted flux of relic neutrinos}

\begin{figure}[t!]
    \centering
    \includegraphics[width=0.5\textwidth]{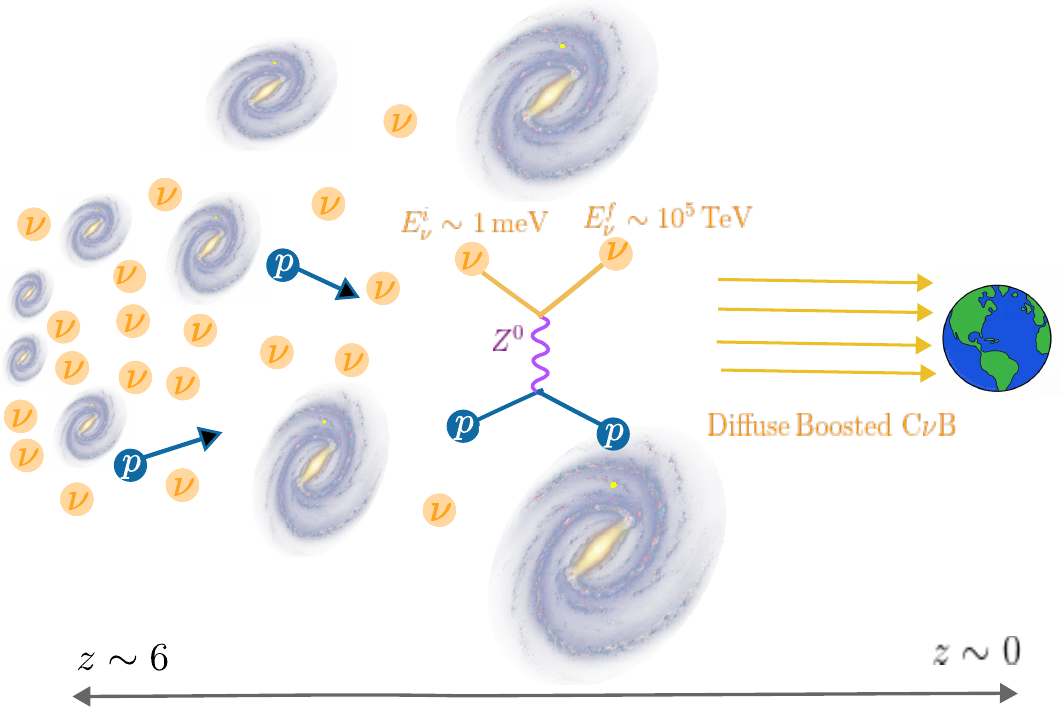}
    \caption{Schematic showing how energetic cosmic-ray protons from galaxies at different redshifts scatter off the C$\nu$B, yielding a diffuse flux boosted neutrinos on Earth. Cosmic rays can transfer a large amount of energy to the C$\nu$B, such that a small fraction of it reaches energies probed by experiments of high-energy neutrinos like IceCube.
    }
    \label{fig:diagram}
\end{figure}

Our setup is illustrated in Fig.~\ref{fig:diagram}. We consider cosmic-rays accelerated in extragalactic sources, which boost the C$\nu$B both within and between galaxies. This is because cosmic rays accelerated to sufficiently high energies will escape the magnetic confinement of their host galaxies \cite{AlvesBatista:2019tlv}, filling the intergalactic medium, which can induce a diffuse flux of boosted relic neutrinos accumulated beyond galactic scales. 

In order to estimate the diffuse flux of boosted C$\nu$B, we assume that the cosmic rays are composed of pure protons; we discuss potential advantages of including nuclei later. The diffuse differential flux of relic neutrinos boosted by cosmic rays at different redshifts to an energy $T_{\nu}^{'}$, redshifted on Earth to an energy $T_{\nu}=T_{\nu}^{'}(1+z)^{-1}$, is given by the following double integral over redshift and proton energy (see, \textit{e.g.,} Refs.~\cite{Ciscar-Monsalvatje:2024tvm, Bringmann:2018cvk} for the analogous formula in the $z \rightarrow 0$ limit),
\begin{widetext}
\begin{equation}
\frac{d \phi_{\nu}}{d T_{\nu}}= \int_{z_{\text {min }}}^{z_{\text {max }}}dz \frac{c}{H_0} \frac{1}{\sqrt{(1+z)^3 \Omega_m+\Omega_{\Lambda}}} f_i(z) n_v(1+z)^3\int_0^{\infty} d T_p \sigma_{p\nu}(T_p)\frac{d \phi_{p}}{d T_p}  \frac{1}{T_\nu^{\max }\left(T_p\right)} \Theta\left[T_\nu^{\max }\left(T_p\right)-T_\nu(1+z)\right],
\label{eq:flux}
\end{equation}
\end{widetext}
where $T_p$ is the kinetic energy of the proton, $f_i(z)$ is the redshift evolution of the cosmic-ray flux, $n_{\nu}=336$ cm$^{-3}$ is the average density of relic neutrinos in the Universe today which we evolve in redshift with the standard $(1+z)^3$, $\sigma_{p\nu}$ is the cosmic-ray neutrino scattering cross section, $d\phi_p/dT_p$ is the cosmic-ray spectrum, and $T_\nu^{\max }$ is the maximum kinetic energy transferred to a neutrino in a single collision. Throughout, we adopt $H_0=67.4 \mathrm{~km} / \mathrm{sec} / \mathrm{Mpc}, \Omega_{\Lambda}=0.685$, and $\Omega_m=0.315$ \cite{Planck:2018vyg}.  Since cosmic-ray protons have much larger energies than the relic neutrinos, which are effectively at rest, the maximum kinetic energy transferred to a neutrino in a single collision is \cite{Ciscar-Monsalvatje:2024tvm, Bringmann:2018cvk, Cappiello:2018hsu, Yin:2018yjn, Khlopov:1987bh}, 
\begin{equation}\label{eq:Tmax}
T_\nu^{\max }=\frac{T_p^2+2 m_p T_p}{T_p+\left(m_p+m_\nu\right)^2 /\left(2 m_\nu\right)},
\end{equation}
where $m_p$ and $m_\nu$ are the proton and neutrino masses, respectively.

Since the source(s) of cosmic rays at the highest energies are still uncertain (see, e.g., \cite{AlvesBatista:2019tlv}), we follow previous studies and consider several source classes. Accordingly, we model the cosmic-ray source redshift evolution as, 
\begin{equation}
f_i(z) = \frac{N_i \left(z\right)}{N_i\left(z_{\rm min}\right)},
\end{equation}
where $N_i(z)$ is the distribution function that determines the normalization of the cosmic-ray flux at different redshifts for three different well-studied distributions: the cosmic star formation rate (SFR), the Fanaroff–Riley II or quasar (QSO) distribution, and the gamma-ray burst (GRB) distribution. These have been obtained from a combination of observations and theoretical inputs, and can be expressed analytically (e.g., \cite{Kotera_2010}). For the normalizations we adopt $z_{\rm min}=2.37 \times 10^{-6}$ which equates approximately to the size of the Milky Way galaxy, $ \simeq 10$ kpc. 

For the SFR evolution, we adopt the result from \cite{Hopkins:2006bw},
\begin{equation}
N_{\mathrm{SFR}}(z)=\frac{1+\left(a_2 z / a_1\right)}{1+\left(z / a_3\right)^{a_4}},
\end{equation}
where $a_1=0.015, a_2=0.10, a_3=3.4$, and $a_4=5.5$. For the QSO distribution, we consider the source density from Ref.~\cite{2002A&A...386...97J}
\begin{equation}
N_{\rm QSO}(z)=-a_0+a_1 z-a_2 z^2+a_3 z^3-a_4 z^4
\end{equation}
with $a_0$=12.49, $a_1$=2.704, $a_2$=1.145, $a_3$=0.1796, and $a_4$=0.01019.
For the GRB distribution function, we use the parametrization from \cite{Le:2006pt},
\begin{equation}
N_{\rm GRB}(z)=\frac{1+a_1}{(1+z)^{-a_2}+a_1(1+z)^{a_3}},
\end{equation}
with the lower (conservative) values corresponding to $a_1=0.005$, $a_2=3.3$, and $a_3=3.0$, and the upper (aggressive) values corresponding to $a_1=0.0001$, $a_2=4$, and $a_3=3.0$. Note that the parameters $a_i$ have errors based on the data to which the functional forms are fitted, but we neglect them in our study, instead capturing the uncertainty by considering different sources. For all sources, we nominally adopt $z_{\rm max}=6$. 

The cosmic-ray flux ${d \phi_{p}}/{d T_p}$ beyond those of the Earth's vicinity cannot be directly measured. For simplicity, we adopt a broken power-law spectrum. For energies below $E_{p}=10^{7}$ GeV we adopt a slope of $-2.7$ which is consistent cosmic rays measured on Earth. Above $E_{p}=10^{7}$ GeV, we adopt power-law spectra $\propto E^{-\alpha}$ where following Ref.~\cite{Kotera_2010} the slope $\alpha$ is source dependent.  The slope $\alpha$ is determined by requiring the cosmic-ray spectrum after propagation through the cosmic microwave background (CMB) to match the observed spectrum on Earth. Due to the effects of energy redshifting, this results in $\alpha$ depending on the cosmic-ray injection redshift distribution. For the star forming rate $\alpha=2.5$, for the quasar distribution $\alpha=2.3$, and for the GRB distribution $\alpha=2.4$ (see Ref.~\cite{Kotera_2010} for details). We later discuss the impact if the slope is not modified.

\begin{figure*}[t!]
    \centering
    \includegraphics[width=0.49\textwidth]{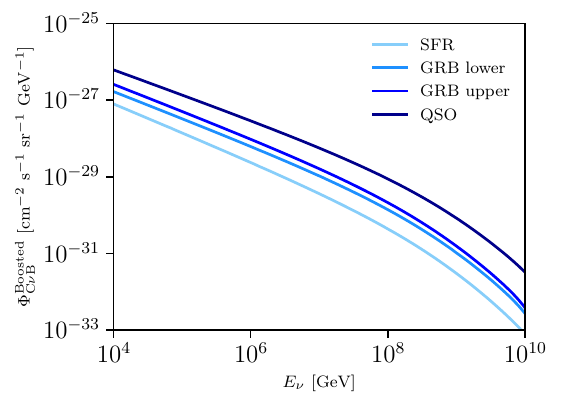}
    \includegraphics[width=0.49\textwidth]{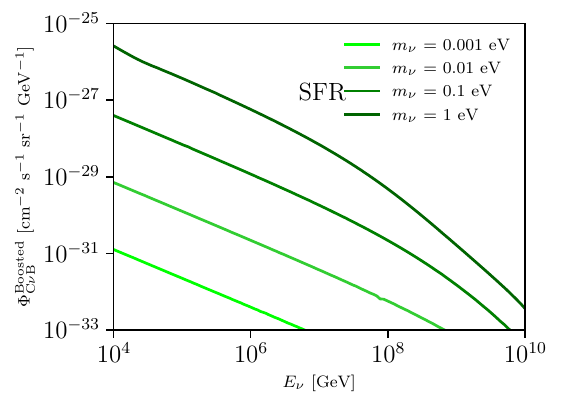}
    \includegraphics[width=0.49\textwidth]{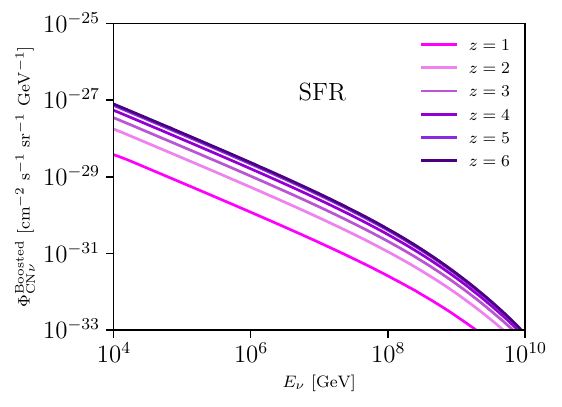}
    \includegraphics[width=0.49\textwidth]{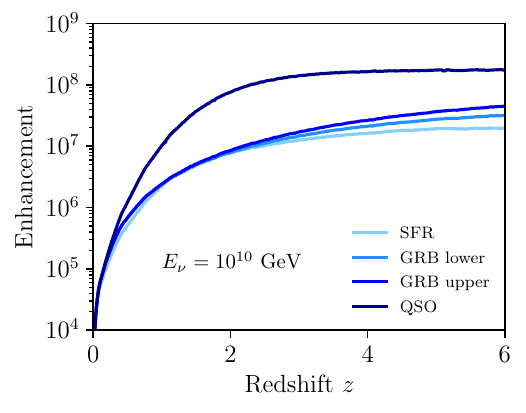}
    \caption{ \textit{Upper left:} Diffuse flux of boosted C$\nu$B from redshift $z=6$ until today, for different dependencies of the highest energy cosmic-ray flux with redshift: star forming rate (SFR), gamma ray burst (GRB upper and GRB lower) and quasars (QSO). The neurino mass is taken to be $m_{\nu}=0.1$ eV. \textit{Upper right:} Diffuse flux of boosted C$\nu$B for different values of the neutrino mass, assuming the cosmic-ray flux normalization and spectra follows the SFR (see main text for details). \textit{Lower left: } Diffuse flux of boosted C$\nu$B for different maximal redshift values considered, $z_{\rm max}$ in Eq.~(\ref{eq:flux}), for a cosmic ray-flux following the SFR and a neutrino mass of $m_{\nu}=0.1$ eV. \textit{Lower right:} Enhancement of the boosted C$\nu$B flux as a function of redshift, with respect to redshift $z=0$, at a neutrino energy of $10^{10}$ GeV.}
    \label{fig:redshift_enhancement}
\end{figure*}

\begin{figure}[t!]
    \centering
    \includegraphics[width=0.49\textwidth]{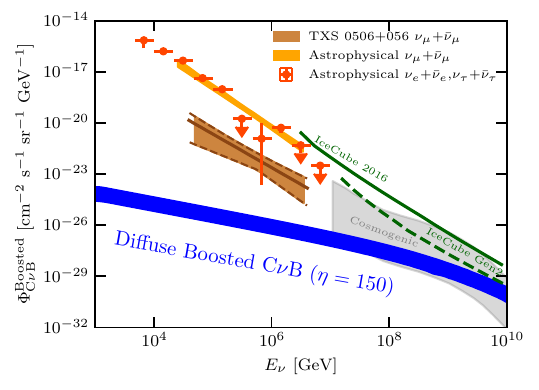}
    \caption{Zoomed diffuse boosted relic neutrino flux at ultra high energies, for an overdensity of relic neutrinos of $\eta=150$ with respect to the average cosmological value.}
    \label{fig:cut_off_overdensity}
\end{figure}

Finally, the neutral current neutrino-proton cross section in the regime of interest of this work ($2 m_\nu E_p<m_p^2$) scales with energy as \cite{Ciscar-Monsalvatje:2024tvm, Formaggio:2012cpf},
\begin{equation}
    \sigma_{p\nu}\sim \frac{G_{F}^{2} E_p^2 m_\nu^2}{\pi m_p^2} 
\end{equation}
where $E_p = T_p + m_p$. This expression for the cross section is valid up to $s \simeq 1 $ GeV$^2$, and we have checked that it agrees with the experimentally measured value within 20$\%$ precision\footnote{An exact analytical expression for the neutrino-proton scattering cross section in the quasi-elastic regime can be found in \cite{PhysRevD.35.785}.}. We proceed with this approximation since the deviation is much smaller than the level of uncertainty introduced by the cosmic-ray flux dependence with redshift or the value of neutrino masses which are many order of magnitudes. 

With the above ingredients, we first check that in the limit of $z_{\rm max} \rightarrow 0$, we reproduce the flux estimate of the Milky Way boosted cosmic neutrino flux from Ref.~\cite{Ciscar-Monsalvatje:2024tvm} with $10\%$ accuracy. This is shown in Fig.~\ref{fig:boosted_flux} in light blue and labeled. 

We next proceed to calculate the diffuse flux of boosted C$\nu$B under the different models of cosmic-ray normalization and redshift evolution. Our main results are summarized in the blue band in Fig.~\ref{fig:boosted_flux}, where we show our predicted flux of boosted relic neutrinos on Earth. The band captures the uncertainty stemming from the different cosmic-ray evolution models and spectra, namely QSO (aggresive) and SFR (conservative). We consider the QSO model aggressive, in part due to it having the largest enhancement (see discussion below), but more because the QSO model is constrained by the null detection of cosmogenic neutrinos, although this conclusion depends on the cosmic-ray maximum energy and cosmic-ray composition \cite{PierreAuger:2019ens}. Overall, the uncertainty arising from the models considered is $\sim$ 1 order of magnitude at most.

Although our predicted boosted cosmic neutrino background flux is below current measurements and sensitivities, it is remarkable that the upper end of our predicted flux exceeds the pessimistic cosmogenic neutrino flux from \cite{GRAND:2018iaj} at energies above $ E_{\nu}\sim 5 \times 10^9$ GeV. More specifically, this situation is realized for QSO model compared to the lower-end of cosmogenic neutrino predictions from Ref.~\cite{GRAND:2018iaj}. 

In the upper panels of Fig.~\ref{fig:redshift_enhancement}, we show how the boosted C$\nu$B flux depends on the inputs. From the top left panel, it can be appreciated that the GRB model lies in between the QSO and SFR models, which effectively bracket the uncertainty of our predicted boosted flux. Furthermore, we show in the upper right panel the dependence of the boosted cosmic neutrino background flux on the neutrino mass, for values in the range $m_{\nu}=[0.001,1]$ eV. The differences can be significant depending on the energy of the boosted neutrino. In the range between $E_{\nu} \sim 10^{7}-10^{9}$ GeV, we find the differences in the spectra to be overall constant, of 1 order of magnitude at most. At lower neutrino energies however, the boosted fluxes increase with lower neutrino masses. For example, at an energy of $E_{\nu}=10^5$ GeV, the boosted flux of an $m_{\nu}=0.01$ eV neutrino is $\sim 2$ orders of magnitude larger than the flux of an $m_{\nu}=1$ eV neutrino. On the contrary, at high energies, the boosted neutrino flux experiences a cut-off resulting from a combination of the cosmic-ray cut-off energy and center of mass of the scattering $s$, which decreases linearly with the neutrino mass. Thus, for sufficiently lighter neutrinos, the cut-off occurs earlier, and the boosted relic neutrino flux can decrease significantly at energies above $E_{\nu} \gtrsim 10^{10}$ GeV.

In the bottom panels of Fig.~\ref{fig:redshift_enhancement}, we explore the dependence on the maximal redshift, $z_{\rm max}$. In the lower left panel, it can be appreciated how the redshifting of neutrino energies after boosting, $T_{\nu}=T_{\nu}^{\prime}(1+z)^{-1}$, convoluted with the source redshift distribution, induces bump-like features in the spectrum. In addition, the majority of boosted relic neutrinos cluster after redshifting at low energies, as expected. We also find that the boosted flux saturates at around $z \sim 5$, with higher-redshifts providing negligible additional contributions. This can be more easily observed in the lower right panel of Fig.~\ref{fig:redshift_enhancement}, where we show the enhancement on the boosted relic neutrino fluxes as a function of redshift. Here, the enhancement is defined as the ratio of the flux of boosted neutrinos at a neutrino energy of $10^{10}$ GeV with respect to the result for redshift $z=0$. The largest enhancement occurs for the QSO model, reaching values as large as $ \sim 2 \times 10^8$, while for the SFR model, the enhancement is roughly one order of magnitude smaller, of $2 \times 10^7$. For these two models, the enhancement saturates by $z \sim 5$. For the QSO model, the saturation occurs closer to redshift $z \sim 3$. It should be noted that for both GRB models under consideration, the enhancement saturates at redshifts larger than $z \sim 6$, due to the increase in source distribution with redshift beyond $z \sim 6$.

The cosmic-ray cut-off energy is a crucial variable. Its exact value is uncertain, in part due to the uncertain source(s) of the highest-energy cosmic rays. As a result, previous studies of cosmic-ray propagation have explored various maximum cut-off energies \cite{Kotera_2010}. The cut-off affects directly the cut-off of the boosted neutrino spectra, which also has an impact for detection prospects with cosmic-ray and neutrino experiments like IceCube, ANITA \cite{ANITA:2018vwl} and Pierre Auger \cite{PierreAuger:2019ens}. 
We caution that in this work we have not performed a full propagation calculation. Since the mean free path of cosmic rays to boost a cosmic background neutrino is longer than the mean free path for photopion production on a CMB photon, a full treatment of cosmic-ray propagation should degrade the cosmic-ray spectrum (the GZK effect) before much boosting of the cosmic neutrino background occurs. For a similar reason, our adopted cosmic-ray slope $\alpha$ above $E_p = 10^7$ GeV will eventually be degraded during propagation. If instead of $\alpha$ we conservatively adopt the observed $-2.7$, the number of the highest energy cosmic rays and thus our predicted boosted relic neutrino flux are reduced, but only by a factor $\sim 10$ (SFR) to $\sim 100$ (QSO). 

So far, we have assumed that the C$\nu$B encountered by cosmic rays can be modeled with the average density in the Universe. However, it is important to emphasize that the average density of the C$\nu$B on scales of galaxies and clusters of galaxies is likely to be significantly enhanced. Gravitational clustering can enhance the neutrino density in these regions at all redshifts \cite{Ringwald:2004np,Hotinli:2023scz}, although particularly at low redshifts when the most massive structures have collapsed, which would enhance the boosted relic neutrino flux above the ones derived here. For illustration, in Fig.~\ref{fig:cut_off_overdensity} we show the boosted relic neutrino flux assuming an overdensity on cosmological distances of $\eta=150$, which would make the cosmic neutrino background to approximately yield the critical density of the Universe \cite{Ciscar-Monsalvatje:2024tvm}. In presence of large overdensities, the neutrino momenta may be larger than its mass. This could only have a sizable effect on the boosted fluxes for overdensities larger than $\eta \gtrsim 10^4$. We estimated the neutrino momenta from the Fermi sphere $n_{\nu}=p_{\nu}^{3}/6\pi^{2}$, obtaining a typical momenta of $m_{\nu} \sim $ meV, smaller than the rest mass of the neutrino $m_{\nu}=0.1$ eV considered. Promisingly, we find that an overdensity of $\eta=150$ could be probed with the next generation IceCube Gen2 experiment, in particular for the QSO model.

It should be noted that in the absence of overdensities, the scattering of cosmic rays off relic neutrinos over cosmological distances occurs in the optically thin limit, i.e., $\tau \sim n_{\nu}\sigma_{p\nu} L_{H} \ll 1$, where $L_H$ denotes the Hubble length. In this regime the number of scattered cosmic rays is proportional to $\tau$. Therefore, most cosmic rays would not lead to a boosted relic neutrino. However, in presence of large neutrino overdensities 
, we enter the regime in between the optically thin and optically thick limits, i.e $\tau \gtrsim 1$, where the scattering probability goes as $\sim 1-e^{-\tau}$, and a significant fraction of cosmic rays could boost relic neutrinos.

Finally, we emphasize that although in our predictions the diffuse flux of boosted relic neutrinos is weaker than the cosmogenic neutrino flux from photopion production on the CMB, the spectra are quite different, which could be exploited in the future to facilitate a differentiation. For example, the cosmogenic neutrino flux presents two broad two ``bumps'' arising from the distribution of two targets: infrared photons vs the CMB photons. On the contrary, the diffuse flux of boosted relic neutrinos made in this study is roughly homogeneous in the energy region comparable to cosmogenic neutrino energies, except for a small bump at small energies arising from neutrino redshifting effects. Furthermore, the cosmogenic neutrino spectra is somewhat harder than our predicted flux.

\section{Discussions and Conclusions}

We have estimated the diffuse flux of the cosmic neutrino background boosted by cosmic-ray protons over the history of the Universe. A fraction of the C$\nu$B can be boosted to sufficiently high energies that afford high-energy neutrino detectors the opportunity to detect the C$\nu$B. The boosted C$\nu$B signal depends on the cosmic-ray flux spectral index and dependence with redshift. We thus assessed multiple scenarios for the redshift distribution of cosmic-ray sources and corresponding spectral indices, namely the SFR, GRB, and QSO distribution functions. We find that the variations are within $\sim 1$ order of magnitude over the energies of interest. 

Intriguingly, the predicted flux of boosted relic neutrinos is several orders of magnitude larger than other boosted relic neutrino signals explored in the literature \cite{Ciscar-Monsalvatje:2024tvm, Franklin:2024enc, DeMarchi:2024zer}. In fact, current limits on cosmogenic neutrinos by IceCube lie only $\sim 4$ orders of magnitude above the predicted diffuse flux of boosted relic neutrinos. With the projected sensitivity of IceCube-Gen2 \cite{IceCube-Gen2:2021rkf}, the gap can reduce to $\sim 3$ orders of magnitude in the QSO model. In this context, relic neutrino overdensities can be generated within galaxies, where cosmic rays are produced, due to gravitational clustering. While we have nominally not assumed any overdensity, some simulations indicate that overdensities of $10^2$ to $10^3$ are expected in galaxies and clusters \cite{Ringwald:2004np, Hotinli:2023scz, Zimmer:2023jbb, Holm:2024zpr}. This indicates that next generation high-energy neutrino telescopes may begin to constrain $\mathcal{O}(1-10)$ overdensities on the cosmic neutrino background over cosmological distances, potentially reaching the average value predicted in $\Lambda$CDM.

The predicted boosted C$\nu$B flux is lower than the cosmogenic neutrino flux arising from interactions of the same cosmic rays with the CMB. The ratio is approximately the ratio of their associated cross sections, since the former depend on the C$\nu$B and the latter the CMB and their number densities are similar. While the cosmogenic neutrinos is the primary neutrino background for the boosted C$\nu$B signal, the two signals have different spectral features which may provide a way to differentiate them in the future. Interestingly, the spectrum of the boosted C$\nu$B also varies significantly for different values of the neutrino mass, which may provide an independent method to determine the neutrino mass in the future. 

There are various ways the sensitivity to boosted relic neutrinos may be further enhanced. For example, we consider only neutral current neutrino-proton interactions. Charged current interaction processes between the cosmic ray and the relic neutrinos can also induce a flux of high-energy neutrinos via boosted muon decay, but we conservatively neglected this contribution. The additional charged current interaction contribution to the diffuse flux of cosmic ray boosted relic neutrinos would peak at somewhat lower energies than the neutral current contribution studied here, but would offer additional boosted signal. Also, besides cosmic-ray protons, cosmic-ray electrons and heavy nuclei could contribute further to boost relic neutrinos. We expect the contribution induced by cosmic-ray electrons to be small, due to their lower fluxes and energies. As for heavy nuclei, these could become dominant at ultra-high energies, where the sensitivity prospects for detection are perhaps the most promising. Heavy nuclei would boost neutrinos with the same cross section than protons, although with an incoherent enhancement proportional to the number of nucleons. Form factors at sufficiently high energies would then needed to be taken into account, which may suppress the cross section for heavy nuclei. However, a larger heavy nuclei composition of ultra high energy cosmic rays would reduce the cosmogenic neutrino flux substantially, which may enhance detection prospects of the diffuse boosted C$\nu$B. 

To conclude, we have shown that the diffuse cosmic-ray boosted C$\nu$B offers a significantly more promising method to detect the C$\nu$B than previous proposals. Paradoxically, the least energetic neutrinos in the Universe may be detected with experiments sensitive to the most energetic neutrinos in the Universe. Such a detection of the long-sought C$\nu$B would be ground-breaking, confirming a cornerstone of modern cosmology, and potentially revealing unknown properties of neutrinos.

\section{Acknowledgments}
We are grateful to Sean Heston, Mar Císcar-Monsalvatje, Ian M.~Shoemaker, John Beacom, and Kohta Murase for useful discussions. GH is supported by the the U.S. Department of Energy under the award number DE-SC0020250 and DE-SC0020262. The work of SH is supported by the U.S.~Department of Energy Office of Science under award number DE-SC0020262, NSF Grant No.~AST1908960 and No.~PHY-2209420, and JSPS KAKENHI Grant Number JP22K03630 and JP23H04899. This work was supported by World Premier International Research Center Initiative (WPI Initiative), MEXT, Japan.

\bibliography{References}

\begin{thebibliography}{35}%
\makeatletter
\providecommand \@ifxundefined [1]{%
 \@ifx{#1\undefined}
}%
\providecommand \@ifnum [1]{%
 \ifnum #1\expandafter \@firstoftwo
 \else \expandafter \@secondoftwo
 \fi
}%
\providecommand \@ifx [1]{%
 \ifx #1\expandafter \@firstoftwo
 \else \expandafter \@secondoftwo
 \fi
}%
\providecommand \natexlab [1]{#1}%
\providecommand \enquote  [1]{``#1''}%
\providecommand \bibnamefont  [1]{#1}%
\providecommand \bibfnamefont [1]{#1}%
\providecommand \citenamefont [1]{#1}%
\providecommand \href@noop [0]{\@secondoftwo}%
\providecommand \href [0]{\begingroup \@sanitize@url \@href}%
\providecommand \@href[1]{\@@startlink{#1}\@@href}%
\providecommand \@@href[1]{\endgroup#1\@@endlink}%
\providecommand \@sanitize@url [0]{\catcode `\\12\catcode `\$12\catcode `\&12\catcode `\#12\catcode `\^12\catcode `\_12\catcode `\%12\relax}%
\providecommand \@@startlink[1]{}%
\providecommand \@@endlink[0]{}%
\providecommand \url  [0]{\begingroup\@sanitize@url \@url }%
\providecommand \@url [1]{\endgroup\@href {#1}{\urlprefix }}%
\providecommand \urlprefix  [0]{URL }%
\providecommand \Eprint [0]{\href }%
\providecommand \doibase [0]{http://dx.doi.org/}%
\providecommand \selectlanguage [0]{\@gobble}%
\providecommand \bibinfo  [0]{\@secondoftwo}%
\providecommand \bibfield  [0]{\@secondoftwo}%
\providecommand \translation [1]{[#1]}%
\providecommand \BibitemOpen [0]{}%
\providecommand \bibitemStop [0]{}%
\providecommand \bibitemNoStop [0]{.\EOS\space}%
\providecommand \EOS [0]{\spacefactor3000\relax}%
\providecommand \BibitemShut  [1]{\csname bibitem#1\endcsname}%
\let\auto@bib@innerbib\@empty
\bibitem [{\citenamefont {Dolgov}\ \emph {et~al.}(1997)\citenamefont {Dolgov}, \citenamefont {Hansen},\ and\ \citenamefont {Semikoz}}]{Dolgov:1997mb}%
  \BibitemOpen
  \bibfield  {author} {\bibinfo {author} {\bibfnamefont {A.~D.}\ \bibnamefont {Dolgov}}, \bibinfo {author} {\bibfnamefont {S.~H.}\ \bibnamefont {Hansen}}, \ and\ \bibinfo {author} {\bibfnamefont {D.~V.}\ \bibnamefont {Semikoz}},\ }\href {\doibase 10.1016/S0550-3213(97)00479-3} {\bibfield  {journal} {\bibinfo  {journal} {Nucl. Phys. B}\ }\textbf {\bibinfo {volume} {503}},\ \bibinfo {pages} {426} (\bibinfo {year} {1997})},\ \Eprint {http://arxiv.org/abs/hep-ph/9703315} {arXiv:hep-ph/9703315} \BibitemShut {NoStop}%
\bibitem [{\citenamefont {Mangano}\ \emph {et~al.}(2005)\citenamefont {Mangano}, \citenamefont {Miele}, \citenamefont {Pastor}, \citenamefont {Pinto}, \citenamefont {Pisanti},\ and\ \citenamefont {Serpico}}]{Mangano:2005cc}%
  \BibitemOpen
  \bibfield  {author} {\bibinfo {author} {\bibfnamefont {G.}~\bibnamefont {Mangano}}, \bibinfo {author} {\bibfnamefont {G.}~\bibnamefont {Miele}}, \bibinfo {author} {\bibfnamefont {S.}~\bibnamefont {Pastor}}, \bibinfo {author} {\bibfnamefont {T.}~\bibnamefont {Pinto}}, \bibinfo {author} {\bibfnamefont {O.}~\bibnamefont {Pisanti}}, \ and\ \bibinfo {author} {\bibfnamefont {P.~D.}\ \bibnamefont {Serpico}},\ }\href {\doibase 10.1016/j.nuclphysb.2005.09.041} {\bibfield  {journal} {\bibinfo  {journal} {Nucl. Phys. B}\ }\textbf {\bibinfo {volume} {729}},\ \bibinfo {pages} {221} (\bibinfo {year} {2005})},\ \Eprint {http://arxiv.org/abs/hep-ph/0506164} {arXiv:hep-ph/0506164} \BibitemShut {NoStop}%
\bibitem [{\citenamefont {Vitagliano}\ \emph {et~al.}(2020)\citenamefont {Vitagliano}, \citenamefont {Tamborra},\ and\ \citenamefont {Raffelt}}]{Vitagliano:2019yzm}%
  \BibitemOpen
  \bibfield  {author} {\bibinfo {author} {\bibfnamefont {E.}~\bibnamefont {Vitagliano}}, \bibinfo {author} {\bibfnamefont {I.}~\bibnamefont {Tamborra}}, \ and\ \bibinfo {author} {\bibfnamefont {G.}~\bibnamefont {Raffelt}},\ }\href {\doibase 10.1103/RevModPhys.92.045006} {\bibfield  {journal} {\bibinfo  {journal} {Rev. Mod. Phys.}\ }\textbf {\bibinfo {volume} {92}},\ \bibinfo {pages} {45006} (\bibinfo {year} {2020})},\ \Eprint {http://arxiv.org/abs/1910.11878} {arXiv:1910.11878 [astro-ph.HE]} \BibitemShut {NoStop}%
\bibitem [{\citenamefont {Aker}\ \emph {et~al.}(2022)\citenamefont {Aker} \emph {et~al.}}]{KATRIN:2021uub}%
  \BibitemOpen
  \bibfield  {author} {\bibinfo {author} {\bibfnamefont {M.}~\bibnamefont {Aker}} \emph {et~al.} (\bibinfo {collaboration} {KATRIN}),\ }\href {\doibase 10.1038/s41567-021-01463-1} {\bibfield  {journal} {\bibinfo  {journal} {Nature Phys.}\ }\textbf {\bibinfo {volume} {18}},\ \bibinfo {pages} {160} (\bibinfo {year} {2022})},\ \Eprint {http://arxiv.org/abs/2105.08533} {arXiv:2105.08533 [hep-ex]} \BibitemShut {NoStop}%
\bibitem [{\citenamefont {Hara}\ and\ \citenamefont {Sato}(1980)}]{hara_sato}%
  \BibitemOpen
  \bibfield  {author} {\bibinfo {author} {\bibfnamefont {T.}~\bibnamefont {Hara}}\ and\ \bibinfo {author} {\bibfnamefont {H.}~\bibnamefont {Sato}},\ }\href {\doibase 10.1143/PTP.64.1089} {\bibfield  {journal} {\bibinfo  {journal} {Progress of Theoretical Physics}\ }\textbf {\bibinfo {volume} {64}},\ \bibinfo {pages} {1089} (\bibinfo {year} {1980})},\ \Eprint {http://arxiv.org/abs/https://academic.oup.com/ptp/article-pdf/64/3/1089/5393707/64-3-1089.pdf} {https://academic.oup.com/ptp/article-pdf/64/3/1089/5393707/64-3-1089.pdf} \BibitemShut {NoStop}%
\bibitem [{\citenamefont {Hara}\ and\ \citenamefont {Sato}(1981)}]{Hara:1980mz}%
  \BibitemOpen
  \bibfield  {author} {\bibinfo {author} {\bibfnamefont {T.}~\bibnamefont {Hara}}\ and\ \bibinfo {author} {\bibfnamefont {H.}~\bibnamefont {Sato}},\ }\href {\doibase 10.1143/PTP.65.477} {\bibfield  {journal} {\bibinfo  {journal} {Prog. Theor. Phys.}\ }\textbf {\bibinfo {volume} {65}},\ \bibinfo {pages} {477} (\bibinfo {year} {1981})}\BibitemShut {NoStop}%
\bibitem [{\citenamefont {C\'\i{}scar-Monsalvatje}\ \emph {et~al.}(2024)\citenamefont {C\'\i{}scar-Monsalvatje}, \citenamefont {Herrera},\ and\ \citenamefont {Shoemaker}}]{Ciscar-Monsalvatje:2024tvm}%
  \BibitemOpen
  \bibfield  {author} {\bibinfo {author} {\bibfnamefont {M.}~\bibnamefont {C\'\i{}scar-Monsalvatje}}, \bibinfo {author} {\bibfnamefont {G.}~\bibnamefont {Herrera}}, \ and\ \bibinfo {author} {\bibfnamefont {I.~M.}\ \bibnamefont {Shoemaker}},\ }\href {\doibase 10.1103/PhysRevD.110.063036} {\bibfield  {journal} {\bibinfo  {journal} {Phys. Rev. D}\ }\textbf {\bibinfo {volume} {110}},\ \bibinfo {pages} {063036} (\bibinfo {year} {2024})},\ \Eprint {http://arxiv.org/abs/2402.00985} {arXiv:2402.00985 [hep-ph]} \BibitemShut {NoStop}%
\bibitem [{\citenamefont {Franklin}\ \emph {et~al.}(2024)\citenamefont {Franklin}, \citenamefont {Martinez-Soler}, \citenamefont {Perez-Gonzalez},\ and\ \citenamefont {Turner}}]{Franklin:2024enc}%
  \BibitemOpen
  \bibfield  {author} {\bibinfo {author} {\bibfnamefont {J.}~\bibnamefont {Franklin}}, \bibinfo {author} {\bibfnamefont {I.}~\bibnamefont {Martinez-Soler}}, \bibinfo {author} {\bibfnamefont {Y.~F.}\ \bibnamefont {Perez-Gonzalez}}, \ and\ \bibinfo {author} {\bibfnamefont {J.}~\bibnamefont {Turner}},\ }\href@noop {} {\  (\bibinfo {year} {2024})},\ \Eprint {http://arxiv.org/abs/2404.02202} {arXiv:2404.02202 [hep-ph]} \BibitemShut {NoStop}%
\bibitem [{\citenamefont {De~Marchi}\ \emph {et~al.}(2024)\citenamefont {De~Marchi}, \citenamefont {Granelli}, \citenamefont {Nava},\ and\ \citenamefont {Sala}}]{DeMarchi:2024zer}%
  \BibitemOpen
  \bibfield  {author} {\bibinfo {author} {\bibfnamefont {A.~G.}\ \bibnamefont {De~Marchi}}, \bibinfo {author} {\bibfnamefont {A.}~\bibnamefont {Granelli}}, \bibinfo {author} {\bibfnamefont {J.}~\bibnamefont {Nava}}, \ and\ \bibinfo {author} {\bibfnamefont {F.}~\bibnamefont {Sala}},\ }\href@noop {} {\  (\bibinfo {year} {2024})},\ \Eprint {http://arxiv.org/abs/2405.04568} {arXiv:2405.04568 [hep-ph]} \BibitemShut {NoStop}%
\bibitem [{\citenamefont {Bauer}\ and\ \citenamefont {Shergold}(2023)}]{Bauer:2022lri}%
  \BibitemOpen
  \bibfield  {author} {\bibinfo {author} {\bibfnamefont {M.}~\bibnamefont {Bauer}}\ and\ \bibinfo {author} {\bibfnamefont {J.~D.}\ \bibnamefont {Shergold}},\ }\href {\doibase 10.1088/1475-7516/2023/01/003} {\bibfield  {journal} {\bibinfo  {journal} {JCAP}\ }\textbf {\bibinfo {volume} {01}},\ \bibinfo {pages} {003} (\bibinfo {year} {2023})},\ \Eprint {http://arxiv.org/abs/2207.12413} {arXiv:2207.12413 [hep-ph]} \BibitemShut {NoStop}%
\bibitem [{\citenamefont {Aartsen}\ \emph {et~al.}(2016{\natexlab{a}})\citenamefont {Aartsen} \emph {et~al.}}]{IceCube:2016uab}%
  \BibitemOpen
  \bibfield  {author} {\bibinfo {author} {\bibfnamefont {M.~G.}\ \bibnamefont {Aartsen}} \emph {et~al.} (\bibinfo {collaboration} {IceCube}),\ }\href {\doibase 10.1103/PhysRevLett.117.241101} {\bibfield  {journal} {\bibinfo  {journal} {Phys. Rev. Lett.}\ }\textbf {\bibinfo {volume} {117}},\ \bibinfo {pages} {241101} (\bibinfo {year} {2016}{\natexlab{a}})},\ \bibinfo {note} {[Erratum: Phys.Rev.Lett. 119, 259902 (2017)]},\ \Eprint {http://arxiv.org/abs/1607.05886} {arXiv:1607.05886 [astro-ph.HE]} \BibitemShut {NoStop}%
\bibitem [{\citenamefont {Richard}\ \emph {et~al.}(2016)\citenamefont {Richard} \emph {et~al.}}]{Super-Kamiokande:2015qek}%
  \BibitemOpen
  \bibfield  {author} {\bibinfo {author} {\bibfnamefont {E.}~\bibnamefont {Richard}} \emph {et~al.} (\bibinfo {collaboration} {Super-Kamiokande}),\ }\href {\doibase 10.1103/PhysRevD.94.052001} {\bibfield  {journal} {\bibinfo  {journal} {Phys. Rev. D}\ }\textbf {\bibinfo {volume} {94}},\ \bibinfo {pages} {052001} (\bibinfo {year} {2016})},\ \Eprint {http://arxiv.org/abs/1510.08127} {arXiv:1510.08127 [hep-ex]} \BibitemShut {NoStop}%
\bibitem [{\citenamefont {Aartsen}\ \emph {et~al.}(2016{\natexlab{b}})\citenamefont {Aartsen} \emph {et~al.}}]{IceCube:2016umi}%
  \BibitemOpen
  \bibfield  {author} {\bibinfo {author} {\bibfnamefont {M.~G.}\ \bibnamefont {Aartsen}} \emph {et~al.} (\bibinfo {collaboration} {IceCube}),\ }\href {\doibase 10.3847/0004-637X/833/1/3} {\bibfield  {journal} {\bibinfo  {journal} {Astrophys. J.}\ }\textbf {\bibinfo {volume} {833}},\ \bibinfo {pages} {3} (\bibinfo {year} {2016}{\natexlab{b}})},\ \Eprint {http://arxiv.org/abs/1607.08006} {arXiv:1607.08006 [astro-ph.HE]} \BibitemShut {NoStop}%
\bibitem [{\citenamefont {Aartsen}\ \emph {et~al.}(2020)\citenamefont {Aartsen} \emph {et~al.}}]{IceCube:2020acn}%
  \BibitemOpen
  \bibfield  {author} {\bibinfo {author} {\bibfnamefont {M.~G.}\ \bibnamefont {Aartsen}} \emph {et~al.} (\bibinfo {collaboration} {IceCube}),\ }\href {\doibase 10.1103/PhysRevLett.125.121104} {\bibfield  {journal} {\bibinfo  {journal} {Phys. Rev. Lett.}\ }\textbf {\bibinfo {volume} {125}},\ \bibinfo {pages} {121104} (\bibinfo {year} {2020})},\ \Eprint {http://arxiv.org/abs/2001.09520} {arXiv:2001.09520 [astro-ph.HE]} \BibitemShut {NoStop}%
\bibitem [{\citenamefont {Abbasi}\ \emph {et~al.}(2021)\citenamefont {Abbasi} \emph {et~al.}}]{IceCube-Gen2:2021rkf}%
  \BibitemOpen
  \bibfield  {author} {\bibinfo {author} {\bibfnamefont {R.}~\bibnamefont {Abbasi}} \emph {et~al.} (\bibinfo {collaboration} {IceCube-Gen2}),\ }\href {\doibase 10.22323/1.395.1183} {\bibfield  {journal} {\bibinfo  {journal} {PoS}\ }\textbf {\bibinfo {volume} {ICRC2021}},\ \bibinfo {pages} {1183} (\bibinfo {year} {2021})},\ \Eprint {http://arxiv.org/abs/2107.08910} {arXiv:2107.08910 [astro-ph.HE]} \BibitemShut {NoStop}%
\bibitem [{\citenamefont {Fang}\ and\ \citenamefont {Murase}(2018)}]{Fang:2017zjf}%
  \BibitemOpen
  \bibfield  {author} {\bibinfo {author} {\bibfnamefont {K.}~\bibnamefont {Fang}}\ and\ \bibinfo {author} {\bibfnamefont {K.}~\bibnamefont {Murase}},\ }\href {\doibase 10.1038/s41567-017-0025-4} {\bibfield  {journal} {\bibinfo  {journal} {Nature Phys.}\ }\textbf {\bibinfo {volume} {14}},\ \bibinfo {pages} {396} (\bibinfo {year} {2018})},\ \Eprint {http://arxiv.org/abs/1704.00015} {arXiv:1704.00015 [astro-ph.HE]} \BibitemShut {NoStop}%
\bibitem [{\citenamefont {\'Alvarez-Mu\~niz}\ \emph {et~al.}(2020)\citenamefont {\'Alvarez-Mu\~niz} \emph {et~al.}}]{GRAND:2018iaj}%
  \BibitemOpen
  \bibfield  {author} {\bibinfo {author} {\bibfnamefont {J.}~\bibnamefont {\'Alvarez-Mu\~niz}} \emph {et~al.} (\bibinfo {collaboration} {GRAND}),\ }\href {\doibase 10.1007/s11433-018-9385-7} {\bibfield  {journal} {\bibinfo  {journal} {Sci. China Phys. Mech. Astron.}\ }\textbf {\bibinfo {volume} {63}},\ \bibinfo {pages} {219501} (\bibinfo {year} {2020})},\ \Eprint {http://arxiv.org/abs/1810.09994} {arXiv:1810.09994 [astro-ph.HE]} \BibitemShut {NoStop}%
\bibitem [{\citenamefont {Alves~Batista}\ \emph {et~al.}(2019)\citenamefont {Alves~Batista} \emph {et~al.}}]{AlvesBatista:2019tlv}%
  \BibitemOpen
  \bibfield  {author} {\bibinfo {author} {\bibfnamefont {R.}~\bibnamefont {Alves~Batista}} \emph {et~al.},\ }\href {\doibase 10.3389/fspas.2019.00023} {\bibfield  {journal} {\bibinfo  {journal} {Front. Astron. Space Sci.}\ }\textbf {\bibinfo {volume} {6}},\ \bibinfo {pages} {23} (\bibinfo {year} {2019})},\ \Eprint {http://arxiv.org/abs/1903.06714} {arXiv:1903.06714 [astro-ph.HE]} \BibitemShut {NoStop}%
\bibitem [{\citenamefont {Bringmann}\ and\ \citenamefont {Pospelov}(2019)}]{Bringmann:2018cvk}%
  \BibitemOpen
  \bibfield  {author} {\bibinfo {author} {\bibfnamefont {T.}~\bibnamefont {Bringmann}}\ and\ \bibinfo {author} {\bibfnamefont {M.}~\bibnamefont {Pospelov}},\ }\href {\doibase 10.1103/PhysRevLett.122.171801} {\bibfield  {journal} {\bibinfo  {journal} {Phys. Rev. Lett.}\ }\textbf {\bibinfo {volume} {122}},\ \bibinfo {pages} {171801} (\bibinfo {year} {2019})},\ \Eprint {http://arxiv.org/abs/1810.10543} {arXiv:1810.10543 [hep-ph]} \BibitemShut {NoStop}%
\bibitem [{\citenamefont {Aghanim}\ \emph {et~al.}(2020)\citenamefont {Aghanim} \emph {et~al.}}]{Planck:2018vyg}%
  \BibitemOpen
  \bibfield  {author} {\bibinfo {author} {\bibfnamefont {N.}~\bibnamefont {Aghanim}} \emph {et~al.} (\bibinfo {collaboration} {Planck}),\ }\href {\doibase 10.1051/0004-6361/201833910} {\bibfield  {journal} {\bibinfo  {journal} {Astron. Astrophys.}\ }\textbf {\bibinfo {volume} {641}},\ \bibinfo {pages} {A6} (\bibinfo {year} {2020})},\ \bibinfo {note} {[Erratum: Astron.Astrophys. 652, C4 (2021)]},\ \Eprint {http://arxiv.org/abs/1807.06209} {arXiv:1807.06209 [astro-ph.CO]} \BibitemShut {NoStop}%
\bibitem [{\citenamefont {Cappiello}\ \emph {et~al.}(2019)\citenamefont {Cappiello}, \citenamefont {Ng},\ and\ \citenamefont {Beacom}}]{Cappiello:2018hsu}%
  \BibitemOpen
  \bibfield  {author} {\bibinfo {author} {\bibfnamefont {C.~V.}\ \bibnamefont {Cappiello}}, \bibinfo {author} {\bibfnamefont {K.~C.~Y.}\ \bibnamefont {Ng}}, \ and\ \bibinfo {author} {\bibfnamefont {J.~F.}\ \bibnamefont {Beacom}},\ }\href {\doibase 10.1103/PhysRevD.99.063004} {\bibfield  {journal} {\bibinfo  {journal} {Phys. Rev. D}\ }\textbf {\bibinfo {volume} {99}},\ \bibinfo {pages} {063004} (\bibinfo {year} {2019})},\ \Eprint {http://arxiv.org/abs/1810.07705} {arXiv:1810.07705 [hep-ph]} \BibitemShut {NoStop}%
\bibitem [{\citenamefont {Yin}(2019)}]{Yin:2018yjn}%
  \BibitemOpen
  \bibfield  {author} {\bibinfo {author} {\bibfnamefont {W.}~\bibnamefont {Yin}},\ }\href {\doibase 10.1051/epjconf/201920804003} {\bibfield  {journal} {\bibinfo  {journal} {EPJ Web Conf.}\ }\textbf {\bibinfo {volume} {208}},\ \bibinfo {pages} {04003} (\bibinfo {year} {2019})},\ \Eprint {http://arxiv.org/abs/1809.08610} {arXiv:1809.08610 [hep-ph]} \BibitemShut {NoStop}%
\bibitem [{\citenamefont {Khlopov}\ and\ \citenamefont {Chechetkin}(1987)}]{Khlopov:1987bh}%
  \BibitemOpen
  \bibfield  {author} {\bibinfo {author} {\bibfnamefont {M.~Y.}\ \bibnamefont {Khlopov}}\ and\ \bibinfo {author} {\bibfnamefont {V.~M.}\ \bibnamefont {Chechetkin}},\ }\href@noop {} {\bibfield  {journal} {\bibinfo  {journal} {Fiz. Elem. Chast. Atom. Yadra}\ }\textbf {\bibinfo {volume} {18}},\ \bibinfo {pages} {627} (\bibinfo {year} {1987})}\BibitemShut {NoStop}%
\bibitem [{\citenamefont {Kotera}\ \emph {et~al.}(2010)\citenamefont {Kotera}, \citenamefont {Allard},\ and\ \citenamefont {Olinto}}]{Kotera_2010}%
  \BibitemOpen
  \bibfield  {author} {\bibinfo {author} {\bibfnamefont {K.}~\bibnamefont {Kotera}}, \bibinfo {author} {\bibfnamefont {D.}~\bibnamefont {Allard}}, \ and\ \bibinfo {author} {\bibfnamefont {A.}~\bibnamefont {Olinto}},\ }\href {\doibase 10.1088/1475-7516/2010/10/013} {\bibfield  {journal} {\bibinfo  {journal} {Journal of Cosmology and Astroparticle Physics}\ }\textbf {\bibinfo {volume} {2010}},\ \bibinfo {pages} {013–013} (\bibinfo {year} {2010})}\BibitemShut {NoStop}%
\bibitem [{\citenamefont {Hopkins}\ and\ \citenamefont {Beacom}(2006)}]{Hopkins:2006bw}%
  \BibitemOpen
  \bibfield  {author} {\bibinfo {author} {\bibfnamefont {A.~M.}\ \bibnamefont {Hopkins}}\ and\ \bibinfo {author} {\bibfnamefont {J.~F.}\ \bibnamefont {Beacom}},\ }\href {\doibase 10.1086/506610} {\bibfield  {journal} {\bibinfo  {journal} {Astrophys. J.}\ }\textbf {\bibinfo {volume} {651}},\ \bibinfo {pages} {142} (\bibinfo {year} {2006})},\ \Eprint {http://arxiv.org/abs/astro-ph/0601463} {arXiv:astro-ph/0601463} \BibitemShut {NoStop}%
\bibitem [{\citenamefont {{Jackson, C. A.}}\ \emph {et~al.}(2002)\citenamefont {{Jackson, C. A.}}, \citenamefont {{Wall, J. V.}}, \citenamefont {{Shaver, P. A.}}, \citenamefont {{Kellermann, K. I.}}, \citenamefont {{Hook, I. M.}},\ and\ \citenamefont {{Hawkins, M. R. S.}}}]{2002A&A...386...97J}%
  \BibitemOpen
  \bibfield  {author} {\bibinfo {author} {\bibnamefont {{Jackson, C. A.}}}, \bibinfo {author} {\bibnamefont {{Wall, J. V.}}}, \bibinfo {author} {\bibnamefont {{Shaver, P. A.}}}, \bibinfo {author} {\bibnamefont {{Kellermann, K. I.}}}, \bibinfo {author} {\bibnamefont {{Hook, I. M.}}}, \ and\ \bibinfo {author} {\bibnamefont {{Hawkins, M. R. S.}}},\ }\href {\doibase 10.1051/0004-6361:20020119} {\bibfield  {journal} {\bibinfo  {journal} {A\&A}\ }\textbf {\bibinfo {volume} {386}},\ \bibinfo {pages} {97} (\bibinfo {year} {2002})}\BibitemShut {NoStop}%
\bibitem [{\citenamefont {Le}\ and\ \citenamefont {Dermer}(2007)}]{Le:2006pt}%
  \BibitemOpen
  \bibfield  {author} {\bibinfo {author} {\bibfnamefont {T.}~\bibnamefont {Le}}\ and\ \bibinfo {author} {\bibfnamefont {C.~D.}\ \bibnamefont {Dermer}},\ }\href {\doibase 10.1086/513460} {\bibfield  {journal} {\bibinfo  {journal} {Astrophys. J.}\ }\textbf {\bibinfo {volume} {661}},\ \bibinfo {pages} {394} (\bibinfo {year} {2007})},\ \Eprint {http://arxiv.org/abs/astro-ph/0610043} {arXiv:astro-ph/0610043} \BibitemShut {NoStop}%
\bibitem [{\citenamefont {Formaggio}\ and\ \citenamefont {Zeller}(2012)}]{Formaggio:2012cpf}%
  \BibitemOpen
  \bibfield  {author} {\bibinfo {author} {\bibfnamefont {J.~A.}\ \bibnamefont {Formaggio}}\ and\ \bibinfo {author} {\bibfnamefont {G.~P.}\ \bibnamefont {Zeller}},\ }\href {\doibase 10.1103/RevModPhys.84.1307} {\bibfield  {journal} {\bibinfo  {journal} {Rev. Mod. Phys.}\ }\textbf {\bibinfo {volume} {84}},\ \bibinfo {pages} {1307} (\bibinfo {year} {2012})},\ \Eprint {http://arxiv.org/abs/1305.7513} {arXiv:1305.7513 [hep-ex]} \BibitemShut {NoStop}%
\bibitem [{\citenamefont {Ahrens}\ \emph {et~al.}(1987)\citenamefont {Ahrens}, \citenamefont {Aronson}, \citenamefont {Connolly}, \citenamefont {Gibbard}, \citenamefont {Murtagh}, \citenamefont {Murtagh}, \citenamefont {Terada}, \citenamefont {White}, \citenamefont {Callas}, \citenamefont {Cutts}, \citenamefont {Hoftun}, \citenamefont {Diwan}, \citenamefont {Lanou}, \citenamefont {Shinkawa}, \citenamefont {Kurihara}, \citenamefont {Amako}, \citenamefont {Kabe}, \citenamefont {Nagashima}, \citenamefont {Suzuki}, \citenamefont {Tatsumi}, \citenamefont {Yamaguchi}, \citenamefont {Abe}, \citenamefont {Beier}, \citenamefont {Doughty}, \citenamefont {Durkin}, \citenamefont {Heagy}, \citenamefont {Hurley}, \citenamefont {Mann}, \citenamefont {Newcomer}, \citenamefont {Williams}, \citenamefont {York}, \citenamefont {Hedin}, \citenamefont {Marx},\ and\ \citenamefont {Stern}}]{PhysRevD.35.785}%
  \BibitemOpen
  \bibfield  {author} {\bibinfo {author} {\bibfnamefont {L.~A.}\ \bibnamefont {Ahrens}}, \bibinfo {author} {\bibfnamefont {S.~H.}\ \bibnamefont {Aronson}}, \bibinfo {author} {\bibfnamefont {P.~L.}\ \bibnamefont {Connolly}}, \bibinfo {author} {\bibfnamefont {B.~G.}\ \bibnamefont {Gibbard}}, \bibinfo {author} {\bibfnamefont {M.~J.}\ \bibnamefont {Murtagh}}, \bibinfo {author} {\bibfnamefont {S.~J.}\ \bibnamefont {Murtagh}}, \bibinfo {author} {\bibfnamefont {S.}~\bibnamefont {Terada}}, \bibinfo {author} {\bibfnamefont {D.~H.}\ \bibnamefont {White}}, \bibinfo {author} {\bibfnamefont {J.~L.}\ \bibnamefont {Callas}}, \bibinfo {author} {\bibfnamefont {D.}~\bibnamefont {Cutts}}, \bibinfo {author} {\bibfnamefont {J.~S.}\ \bibnamefont {Hoftun}}, \bibinfo {author} {\bibfnamefont {M.}~\bibnamefont {Diwan}}, \bibinfo {author} {\bibfnamefont {R.~E.}\ \bibnamefont {Lanou}}, \bibinfo {author} {\bibfnamefont {T.}~\bibnamefont {Shinkawa}}, \bibinfo {author} {\bibfnamefont {Y.}~\bibnamefont {Kurihara}}, \bibinfo {author}
  {\bibfnamefont {K.}~\bibnamefont {Amako}}, \bibinfo {author} {\bibfnamefont {S.}~\bibnamefont {Kabe}}, \bibinfo {author} {\bibfnamefont {Y.}~\bibnamefont {Nagashima}}, \bibinfo {author} {\bibfnamefont {Y.}~\bibnamefont {Suzuki}}, \bibinfo {author} {\bibfnamefont {S.}~\bibnamefont {Tatsumi}}, \bibinfo {author} {\bibfnamefont {Y.}~\bibnamefont {Yamaguchi}}, \bibinfo {author} {\bibfnamefont {K.}~\bibnamefont {Abe}}, \bibinfo {author} {\bibfnamefont {E.~W.}\ \bibnamefont {Beier}}, \bibinfo {author} {\bibfnamefont {D.~C.}\ \bibnamefont {Doughty}}, \bibinfo {author} {\bibfnamefont {L.~S.}\ \bibnamefont {Durkin}}, \bibinfo {author} {\bibfnamefont {S.~M.}\ \bibnamefont {Heagy}}, \bibinfo {author} {\bibfnamefont {M.}~\bibnamefont {Hurley}}, \bibinfo {author} {\bibfnamefont {A.~K.}\ \bibnamefont {Mann}}, \bibinfo {author} {\bibfnamefont {F.~M.}\ \bibnamefont {Newcomer}}, \bibinfo {author} {\bibfnamefont {H.~H.}\ \bibnamefont {Williams}}, \bibinfo {author} {\bibfnamefont {T.}~\bibnamefont {York}}, \bibinfo {author}
  {\bibfnamefont {D.}~\bibnamefont {Hedin}}, \bibinfo {author} {\bibfnamefont {M.~D.}\ \bibnamefont {Marx}}, \ and\ \bibinfo {author} {\bibfnamefont {E.}~\bibnamefont {Stern}},\ }\href {\doibase 10.1103/PhysRevD.35.785} {\bibfield  {journal} {\bibinfo  {journal} {Phys. Rev. D}\ }\textbf {\bibinfo {volume} {35}},\ \bibinfo {pages} {785} (\bibinfo {year} {1987})}\BibitemShut {NoStop}%
\bibitem [{\citenamefont {Aab}\ \emph {et~al.}(2019)\citenamefont {Aab} \emph {et~al.}}]{PierreAuger:2019ens}%
  \BibitemOpen
  \bibfield  {author} {\bibinfo {author} {\bibfnamefont {A.}~\bibnamefont {Aab}} \emph {et~al.} (\bibinfo {collaboration} {Pierre Auger}),\ }\href {\doibase 10.1088/1475-7516/2019/10/022} {\bibfield  {journal} {\bibinfo  {journal} {JCAP}\ }\textbf {\bibinfo {volume} {10}},\ \bibinfo {pages} {022} (\bibinfo {year} {2019})},\ \Eprint {http://arxiv.org/abs/1906.07422} {arXiv:1906.07422 [astro-ph.HE]} \BibitemShut {NoStop}%
\bibitem [{\citenamefont {Gorham}\ \emph {et~al.}(2018)\citenamefont {Gorham} \emph {et~al.}}]{ANITA:2018vwl}%
  \BibitemOpen
  \bibfield  {author} {\bibinfo {author} {\bibfnamefont {P.~W.}\ \bibnamefont {Gorham}} \emph {et~al.} (\bibinfo {collaboration} {ANITA}),\ }\href {\doibase 10.1103/PhysRevD.98.022001} {\bibfield  {journal} {\bibinfo  {journal} {Phys. Rev. D}\ }\textbf {\bibinfo {volume} {98}},\ \bibinfo {pages} {022001} (\bibinfo {year} {2018})},\ \Eprint {http://arxiv.org/abs/1803.02719} {arXiv:1803.02719 [astro-ph.HE]} \BibitemShut {NoStop}%
\bibitem [{\citenamefont {Ringwald}\ and\ \citenamefont {Wong}(2004)}]{Ringwald:2004np}%
  \BibitemOpen
  \bibfield  {author} {\bibinfo {author} {\bibfnamefont {A.}~\bibnamefont {Ringwald}}\ and\ \bibinfo {author} {\bibfnamefont {Y.~Y.~Y.}\ \bibnamefont {Wong}},\ }\href {\doibase 10.1088/1475-7516/2004/12/005} {\bibfield  {journal} {\bibinfo  {journal} {JCAP}\ }\textbf {\bibinfo {volume} {12}},\ \bibinfo {pages} {005} (\bibinfo {year} {2004})},\ \Eprint {http://arxiv.org/abs/hep-ph/0408241} {arXiv:hep-ph/0408241} \BibitemShut {NoStop}%
\bibitem [{\citenamefont {Hotinli}\ \emph {et~al.}(2023)\citenamefont {Hotinli}, \citenamefont {Sabti}, \citenamefont {North},\ and\ \citenamefont {Kamionkowski}}]{Hotinli:2023scz}%
  \BibitemOpen
  \bibfield  {author} {\bibinfo {author} {\bibfnamefont {S.~C.}\ \bibnamefont {Hotinli}}, \bibinfo {author} {\bibfnamefont {N.}~\bibnamefont {Sabti}}, \bibinfo {author} {\bibfnamefont {J.}~\bibnamefont {North}}, \ and\ \bibinfo {author} {\bibfnamefont {M.}~\bibnamefont {Kamionkowski}},\ }\href {\doibase 10.1103/PhysRevD.108.103504} {\bibfield  {journal} {\bibinfo  {journal} {Phys. Rev. D}\ }\textbf {\bibinfo {volume} {108}},\ \bibinfo {pages} {103504} (\bibinfo {year} {2023})},\ \Eprint {http://arxiv.org/abs/2306.15715} {arXiv:2306.15715 [astro-ph.CO]} \BibitemShut {NoStop}%
\bibitem [{\citenamefont {Zimmer}\ \emph {et~al.}(2023)\citenamefont {Zimmer}, \citenamefont {Correa},\ and\ \citenamefont {Ando}}]{Zimmer:2023jbb}%
  \BibitemOpen
  \bibfield  {author} {\bibinfo {author} {\bibfnamefont {F.}~\bibnamefont {Zimmer}}, \bibinfo {author} {\bibfnamefont {C.~A.}\ \bibnamefont {Correa}}, \ and\ \bibinfo {author} {\bibfnamefont {S.}~\bibnamefont {Ando}},\ }\href {\doibase 10.1088/1475-7516/2023/11/038} {\bibfield  {journal} {\bibinfo  {journal} {JCAP}\ }\textbf {\bibinfo {volume} {11}},\ \bibinfo {pages} {038} (\bibinfo {year} {2023})},\ \Eprint {http://arxiv.org/abs/2306.16444} {arXiv:2306.16444 [astro-ph.CO]} \BibitemShut {NoStop}%
\bibitem [{\citenamefont {Holm}\ \emph {et~al.}(2024)\citenamefont {Holm}, \citenamefont {Zentarra},\ and\ \citenamefont {Oldengott}}]{Holm:2024zpr}%
  \BibitemOpen
  \bibfield  {author} {\bibinfo {author} {\bibfnamefont {E.~B.}\ \bibnamefont {Holm}}, \bibinfo {author} {\bibfnamefont {S.}~\bibnamefont {Zentarra}}, \ and\ \bibinfo {author} {\bibfnamefont {I.~M.}\ \bibnamefont {Oldengott}},\ }\href@noop {} {\  (\bibinfo {year} {2024})},\ \Eprint {http://arxiv.org/abs/2404.11295} {arXiv:2404.11295 [hep-ph]} \BibitemShut {NoStop}%
\end{thebibliography}%

\end{document}